\documentclass[reprint,superscriptaddress,preprintnumbers,nofootinbib,nobibnotes,amsmath,amssymb,prl]{revtex4-2}
\pdfoutput=1
\usepackage{amsmath}
\usepackage{amssymb}
\usepackage{CJK}
\usepackage{graphicx}
\usepackage{dcolumn}
\usepackage{bm}
\usepackage[colorlinks=true,pdfstartview=FitV,breaklinks=true]{hyperref}
\usepackage[dvipsnames,table]{xcolor}
\hypersetup{urlcolor=BlueViolet,
	    citecolor=Plum,
	    linkcolor=PineGreen}

\newcommand{\mms}{m_{\mathrm{MS}}}

\begin{document}

\title{Comment on: Cosmological black holes are not described by the Thakurta metric}
\author{C\'eline B\oe hm}
\email{celine.boehm@sydney.edu.au}
\author{Archil Kobakhidze}
\email{archil.kobakhidze@sydney.edu.au}
\author{Ciaran A. J. O'Hare}
\email{ciaran.ohare@sydney.edu.au}
\author{Zachary S. C. Picker}
\email{zachary.picker@sydney.edu.au}
\affiliation{Sydney Consortium for Particle Physics and Cosmology, \\
 School of Physics, The University of Sydney, NSW 2006, Australia }
\author{Mairi Sakellariadou}
\email{mairi.sakellariadou@kcl.ac.uk}
\affiliation{Theoretical Particle Physics and Cosmology Group, Physics Department, King’s College London, University of London, Strand, London WC2R 2LS, UK}

\begin{abstract}
Recently, H\"utsi et al.~\cite{Hutsi:2021nvs} critiqued our work that reconsidered the mathematical description of cosmological black holes. In this short comment, we highlight some of the conceptual issues with this criticism in relation to the interpretation of the quasi-local Misner-Sharp mass, and the fact that our description of cosmological black holes does not impose any assumptions about matter accretion.
\end{abstract}

\maketitle

{\bf Introduction}---To model the behaviour of primordial black holes (PBHs) in the very early Universe, it is necessary to determine a way to embed black hole solutions in an expanding cosmological background. 
This issue was realised early on by Carr and Hawking~\cite{carr10.1093/mnras/168.2.399} in response to Zel'dovich and Novikov~\cite{zeldovich1966AZh....43..758Z} who, several years earlier, had dismissed the existence of PBHs on the grounds that black holes accreting at the Bondi accretion rate would grow far too large. Carr and Hawking pointed out that these black holes needed a proper cosmological embedding, and showed that the accretion rate is negligible when the global geometry of spacetime at large distances is given by the Friedmann-Lema\^{i}tre-Robertson-Walker (FLRW) metric. This work reopened the possibility for black holes to have existed at very early times and paved the way for PBHs to become a viable dark matter candidate. Although Carr and Hawking did not suggest a specific metric, the point stands that the cosmological embedding of black holes is extremely relevant for the phenomenology of PBHs. 

In two recent articles~\cite{Boehm:2020jwd,Picker:2021jxl} we studied the fate of PBHs described in a way that realises the arguments of Carr and Hawking, i.e. by a black hole solution that asymptotically approaches the FLRW metric. We found that the late-time attractor of a class of generalised McVittie metrics --- the Thakurta spacetime~\cite{Thakurta} --- stands out as a particularly appealing example of a metric with this property. In contrast, alternative solutions~\cite{McVittie:1933zz,Einstein:1945id,Lemaitre:1933gd,Tolman:1934za,Bondi:1947fta,Faraoni:2018xwo} face numerous pathological issues.  

The phenomenological consequences of this choice of metric are rather dramatic in both mass ranges we have considered so far. For super-Solar mass PBHs, we found that binaries rarely form in the early Universe, and those that do must form so close together that they coalesce extremely rapidly~\cite{Boehm:2020jwd}, alleviating earlier constraints from LIGO/Virgo~\cite{Sasaki:2016jop,Ali-Haimoud:2016mbv}. On the other hand, asteroid-mass PBHs described by the Thakurta metric are highly unstable and are subsequently disfavoured as a dark matter candidate~\cite{Picker:2021jxl}.

The underlying message behind this work is that the choice of metric to embed black holes in cosmology is highly consequential for PBH dark matter and should not be brushed aside. As such, we encourage critical discussion of this issue in the community, such as the recent work of H\"utsi et al.~\cite{Hutsi:2021nvs}. We believe, however, that the arguments presented in~\cite{Hutsi:2021nvs} are based on a misinterpretation of the general physical picture that we originally presented. We use this opportunity to clarify the situation further.

\vspace{1em}

{\bf Spherically symmetric metrics}.---Before we discuss the specific case of the Thakurta metric, we can look at spacetime metrics with central inhomogeneities in general. We can write any spherically symmetric spacetime in the following Schwarzschild-like form~\cite{Nielsen2006,Abreu:2010ru},
\begin{align}\label{kodamametric}
    \mathrm{d}s^2 = \left(1-\frac{2Gm_{\rm{MS}}}{R}\right)\mathrm{d}\tau^2 &- \left(1-\frac{2Gm_{\rm{MS}}}{R}\right)^{-1}\mathrm{d}R^2 \nonumber\\
    &- R^2 \left(\mathrm{d}\theta^2+\sin^2\theta \,\mathrm{d}\phi^2\right)~,
\end{align}
where $R$ is the areal (or `physical') radius and the time coordinate $\tau$ is defined such that the norm of the Kodama time coincides with the norm of the Kodama vector~\cite{Kodama:1979vn,Abreu:2010ru,Picker:2021jxl}. The quantity playing the role of a mass, $\mms$, is known as the Misner-Sharp mass and is generally a function of $R$ and $\tau$. The Misner-Sharp mass is independent of the coordinate choice, but its physical interpretation is most transparent in the coordinates adopted in Eq.(\ref{kodamametric}). Taking a volume integral of the $00$ component of the Einstein equations we obtain: 
\begin{align}
    \mms = \int_V \mathrm{d}^3x \sqrt{-g}~  T^0_0~,
    \label{mass}
\end{align}
where $T^0_0$ is the energy density component of the total energy-momentum tensor that supports the spacetime of Eq.(\ref{kodamametric}). The Misner-Sharp mass is the total energy within a spherical volume of a given radius $R$ at a time $\tau$. This can be contrasted with the Arnowitt-Deser-Misner (ADM) mass, which is the mass observed by a distant \emph{inertial} observer in asymptotically flat spacetime. While the ADM mass can be attributed entirely to a central inhomogeneity embedded in empty space, the Misner-Sharp mass is a \textit{local} gravitating mass. These two notions of mass are different in general and have different uses that should not be conflated, as we will now discuss. 

\vspace{1em}

{\bf Thakurta black hole description}.---The Thakurta metric~\cite{Thakurta} corresponds to the form shown above with the Misner-Sharp mass given by
\begin{equation}\label{MS}
m_{\mathrm{MS}}=ma(t)+\frac{H^2 R^3}{2Gf(R)}~,
\end{equation}
where $f(R) = 1-2G m a(t)/R$, $a(t)$ is the scale factor, and $H = \dot{a}/a$. This metric is the solution of Einstein's equations with the energy-momentum tensor describing an imperfect fluid at rest with respect to the central inhomogeneity, $\vec{u}=0$, with a radial heat flow $q_{\mu}$:
\begin{align}\label{tensor}
    T_{\mu\nu} &= \left(\rho+P\right)u_\mu u_\nu + g_{\mu\nu}P + q_{(\mu}u_{\nu)}~,\\
    q_{\mu} &= (0,q,0,0)~,\nonumber\\
    u_{\mu} &= (u,0,0,0)~.\nonumber
\end{align}

The mass parameter $m$ in (\ref{MS}) can be thought of as the true physical mass of the black hole in the sense that it is related to the mass of the overdensity from which it formed: $m = \gamma m_H$, where $\gamma \approx 0.2$ accounts for the details of collapse~\cite{Carr:2020gox}, and $m_H$ is the mass within the cosmological horizon. This mass would be the inertial ADM mass in the appropriate static limit. The energy density $\rho$ and pressure $P$ in Eq.(\ref{tensor}) encompass the total contribution from various components of the cosmic fluid (dark and baryonic matter, and radiation) and they dictate the expansion of the Universe at large scales, $R\gg\left(2Gma/H^2\right)^{1/3}$. In this regime, the second term in Eq.(\ref{MS}) dominates and the Misner-Sharp mass becomes the total mass of the cosmic fluid within that radius $R$: $m_{\mathrm{MS}}\approx \frac{H^2 R^3}{2G}=\frac{4\pi}{3}R^3\rho$. In the opposing limit, $R\ll\left(2Gma/H^2\right)^{1/3}$, the first term in Eq.(\ref{MS}) dominates, so the local gravitating mass is given by $ma$.   

The Misner-Sharp mass is useful when we need a local measure of the gravitating mass within some radius, but it should not be used to calculate global quantities related to a cosmological population of black holes. In Ref.~\cite{Hutsi:2021nvs} the authors associate the Misner-Sharp mass in the small-scale limit, $m_{MS}\approx ma$, with the PBH's inertial mass and use it to estimate the cosmological energy density of PBHs from their number density $n$. This leads to the scaling $\rho_{\rm BH}=m_{\mathrm{MS}}n \propto a^{-2}$, in clear violation of the cosmological scaling expected of cold dark matter. However, this is an improper usage of the Misner-Sharp mass. An estimate like this assumes that the energy density of the Universe's dark matter is comprised of PBHs. Yet the Misner-Sharp mass incorporates both the effective mass of the central inhomogeneity, as well as the mass of the surrounding cosmological fluid. Therefore this notion of mass cannot be the appropriate one to use for estimating the global energy density of PBHs, since its definition already contains the very density that we are trying to estimate.

Fortunately, this apparent contradiction is lifted when we use the appropriate definition of mass for the quantity of interest. The PBHs were originally formed from local overdensities of mass $m$, and so the global PBH energy density is related to the number density through $\rho_{\mathrm{BH}} = m n$. The population of PBHs then scales as $\rho_{\rm BH}=mn\propto a^{-3}$, as expected for cold dark matter.

\vspace{1em}

{\bf The Thakurta metric is not eternal}.---Another way to think about Eq.(\ref{MS}) is not in terms of a black hole growing in size, but rather in terms of a black hole that appears lighter than it truly is because the surrounding space is expanding. However, this phenomenon is only a feature of dynamics in the vicinity of a black hole when the effects of the surrounding cosmological expansion are important. As was correctly pointed out in Ref.~\cite{Hutsi:2021nvs}, the Thakurta metric can therefore no longer be valid once the black hole decouples from the Hubble flow in some way. Past the time of decoupling, the black holes `regain' their static physical masses and we can proceed with calculations, such as the binary coalescence time, following entirely standard methods~\cite{Peters:1964zz}. This is what we did in Ref.~\cite{Boehm:2020jwd}. Retaining the growing Misner-Sharp mass for such a calculation would not be correct, as the exercise presented in the appendix of Ref.~\cite{Hutsi:2021nvs} shows---so in this respect we are in agreement.

\vspace{1em}

{\bf No pathologies}.---It is worth stating that we do not claim that the Thakurta metric treatment is definitively the unique way to embed black holes in an expanding background, only that it is the least problematic of the options we have been able to find (including especially the Schwarzschild metric). A valid criticism of the Thakurta metric would therefore be if it were shown that it \emph{did} possess some pathological issues that render its selection over alternatives less secure. For example, in Ref.~\cite{Carrera:2009ve} it was claimed that the Thakurta metric does not escape the spacelike singularity at $R=2Gma$, which is a pathology suffered by the McVittie solution. While there is indeed a spacelike singularity at this radius, the Thakurta apparent horizon is $R_b = \frac{1}{2H}\left(1-\sqrt{1-8HGma}\,\right)>2Gma$, so this singularity is always hidden behind the apparent horizon, thus rendering it harmless in precisely the same way as with the central singularity.

Some further issues arise in Ref.~\cite{Mello:2016irl}, in which the Thakurta metric is studied in detail for different asymptotic scenarios for $a(t)$. Because we assume that the Thakurta metric does not describe the black hole forever, the conclusions of this paper for unbounded scale factors do not apply here.

\vspace{1em}

{\bf No fluid accretion}.---In Ref.~\cite{Hutsi:2021nvs} the authors interpret the growth of the Misner-Sharp mass as a consequence of accretion. The bulk of the criticism that then follows involves arguments that discredit our use of the Thakurta metric on the grounds that the detailed physics of accretion is not properly accounted for.

Accretion involves the flow of matter onto a black hole. However, if we inspect the energy-momentum tensor that sources the Thakurta black hole Eq.(\ref{tensor}) we can see that that the fluid is at rest relative to the black hole, $\vec u = 0$. There is a radial energy flow, described by the $T^r_t$ component, but this does \textit{not} imply any flow of matter, and should be thought of as a temperature gradient towards the black hole. The growing Misner-Sharp mass of the Thakurta metric should not be interpreted as the consequence of fluid accretion, but as a feature of this specific geometry. 

In more general terms, if one did wish to incorporate accretion, this is conventionally described in terms of the inflow rate of matter --- not in terms of the increase of the black hole mass (which should be negligible, otherwise the backreaction onto the surrounding fluid would need to be modelled). Nevertheless, it turns out that accretion is readily achieved using the Thakurta metric by including a radial component in the fluid velocity to model the radial flow of mass.\footnote{We should note that this point can be easily misconstrued when reading Sec.~3.4.2 of the review of black hole embeddings by Faraoni~\cite{Faraoni:2018xwo}. There, the black hole mass growth is erroneously attributed in part to the ``infall of matter'' even though there is no radial mass flow in the energy-momentum tensor being described. The case when there is radial mass flow is discussed directly after this in Sec.~3.4.3.} This strategy was adopted by Babichev et al.~\cite{Babichev:2005py,Babichev:2018ubo} who studied accretion in this way. Our previous studies \cite{Boehm:2020jwd,Picker:2021jxl}  do not involve, nor require, any modelling of accretion. A detailed analysis of accretion onto Thakurta black holes is worthwhile and interesting and will be presented in a forthcoming paper.

\vspace{1em}

{\bf Conclusion}.---The main criticisms of Ref.~\cite{Hutsi:2021nvs} can be addressed with three key points. The first is that the Thakurta metric should only be applied up until the PBHs decouple from the Hubble flow. The second is that the Misner-Sharp mass is only an effective, \textit{local} measure of gravitating mass. The final point is that the stress-energy tensor that sources the Thakurta metric does not assume any matter accretion because the fluid velocity has no radial component. 


\bibliography{bib.bib}
\bibliographystyle{apsrev4-1}

\end{document}